\documentclass[aps,floatfix,prl,superscriptaddress,twocolumn]{revtex4-1}
\pdfoutput=1
\usepackage[hyperindex,breaklinks]{hyperref}
\hypersetup{linkcolor=blue, citecolor=blue, colorlinks=true}

\usepackage{bbm}
\usepackage{graphicx}
\usepackage{dcolumn}
\usepackage{subfigure}
\usepackage{amsmath}
\usepackage{amsfonts}
\usepackage{appendix}
\usepackage{feynmf}
\usepackage{eepic}
\usepackage{lipsum} 
\usepackage{enumerate}  
\usepackage{amssymb} 
\usepackage[english]{babel}  
\usepackage{xcolor}

\newcommand{\SRO}{Sr$_2$RuO$_4$}
\newcommand{\bk}{\mathbf k}

\renewcommand{\vec}[1]{\mathbf{#1}}

\usepackage{times}

\begin{document}
\title{Possibility of mixed helical p-wave pairings in \SRO}
\author{Wen Huang}
\email[]{huangw3@sustech.edu.cn}
\address{Shenzhen Institute for Quantum Science and Engineering \& Guangdong Provincial Key Laboratory of Quantum Science and Engineering, Southern University of Science and Technology, Shenzhen 518055, Guangdong, China}
\author{Zhiqiang Wang}
\email[]{zqwang@uchicago.edu}
\affiliation{James Franck Institute, University of Chicago, Chicago, Illinois 60637, USA}
\date{\today}

\begin{abstract}
The exact nature of unconventional superconductivity in \SRO~remains a mystery. At the phenomenological level, no superconducting order parameter proposed thus far seems able to coherently account for all essential experimental signatures. Among the latter is the prominent polar Kerr effect, which implies a nonzero ac anomalous Hall conductivity. Assuming the Kerr effect is intrinsic, it can be accounted for by a bulk chiral Cooper pairing with nonzero orbital angular momentum, such as $p+ip$ or $d+id$, which, however, has difficulties in being reconciled with other experimental results. Given the situation, in this paper, we propose alternative possibilities with complex mixtures of distinct helical p-wave order parameters, namely, $A_{1u}+iA_{2u}$ and $B_{1u}+iB_{2u}$ in the group theory nomenclature. These states essentially consist of two copies of chiral p-wave pairings with opposite chirality and different pairing amplitudes, and therefore support intrinsic Hall and Kerr effects. We further show that these states exhibit salient features that may explain several other key observations in this material, including the absence of spontaneous edge current, a substantial Knight shift drop, and possibly signatures in uniaxial strain and ultrasound measurements. 
\end{abstract}

\maketitle
The nature of the unconventional superconductivity in \SRO~is a puzzle that remains unsolved~\cite{Maeno:94,Maeno:01,Mackenzie:03,Kallin:09,Kallin:12,Maeno:12,Liu:15,Kallin:16,Mackenzie:17,Leggett:20}. Its normal state properties have been characterized with unprecedented accuracy --- an incredible feat that is not often seen in other unconventional superconductors. The electron correlation in this material is moderate, and superconductivity emerges out of a well-behaved Fermi liquid~\cite{Mackenzie:03}. These have led to the popular perception that identifying its pairing symmetry and resolving its pairing mechanism should be well within the reach of established theories and experimental techniques. Yet, despite tremendous efforts over the past 20-plus years, neither of these two key issues has been settled. This is largely due to the lack of an order parameter that could coherently interpret all the essential experimental observations~\cite{Mackenzie:17,Leggett:20}. 

Arguably, one principal property of the superconductivity in \SRO~is the condensation of at least two superconducting order parameters. This has been inferred from a variety of experiments, including $\mu$SR~\cite{Luke:98,Grinenko:20}, optical polar Kerr effect~\cite{Xia:06}, Josephson interferometry~\cite{Kidwingira:06}, ultrasound~\cite{Ghosh:20,Benhabib:20}, etc. A multi-component pairing is realized if the Cooper pairing is developed in a multi-dimensional irreducible representation (irrep) of the crystal point symmetry group. One notable example is the chiral p-wave pairing, i.e. $(k_x+ik_y)\hat{z}$, which belongs to the $E_u$ irrep of the $D_{4h}$ group. Here, $\hat{z}$ denotes the orientation of the so-called $\vec{d}$-vector, which describes the spin configuration of a spin-triplet pairing~\cite{Mackenzie:03}. Another possibility is coexisting order parameters from distinct one-dimensional irreps. In this scenario, the multiple components typically condense below different critical temperatures except for accidental degeneracy.

\begin{figure}[htp]
\includegraphics[width=8.5cm]{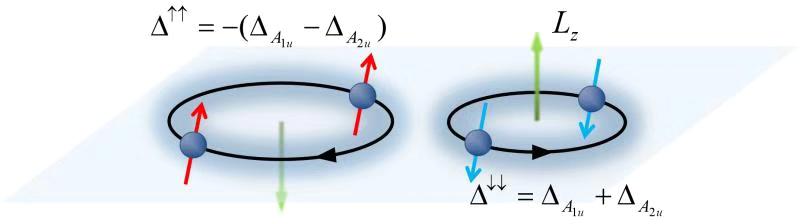}
\caption{(color online) Illustration of the Cooper pairing in the mixed helical p-wave $A_{1u}+iA_{2u}$ state in a single-orbital model. The state consists of two copies of chiral p-wave model with opposite chirality. The Cooper pairs are drawn with different size to reflect their different pairing amplitudes. Green arrows indicate the direction of the Cooper pair orbital angular momentum $L_z$.  Another candidate state $B_{1u}+iB_{2u}$ has a similar pairing configuration. }
\label{figSketch}
\end{figure}

In light of the evidence of time-reversal symmetry breaking from the $\mu$SR and the polar Kerr effect measurements~\cite{Luke:98,Grinenko:20,Xia:06}, the multiple components presumably form a complex pairing, such that the state does not return to itself after a time-reversal operation. The Kerr effect, as well as the closely related anomalous Hall effect, intuitively implies an underlying pairing with finite orbital angular momentum (such as $p+ip$, $d+id$, etc.), as has been substantiated by a recent study~\cite{Zhang:20}. Hence, if the Kerr effect in \SRO~is intrinsic (i.e. not related to extrinsic effects such as impurities or lattice dislocations), its superconducting symmetry is limited to a few candidates. However, as it currently stands, none of the chiral superconducting states appears able to reconcile all of the key observations~\cite{Zhang:20}. For example, although $p_x+ip_y$ or $d_{xz}+id_{yz}$ pairing can naturally explain the Kerr effect, the former is strongly disfavored by the recent NMR measurements~\cite{Pustogow:19,Ishida:20}, and both generally exhibit finite spontaneous surface current that is not detected~\cite{Kirtley:07,Hicks:10}. 

In this work, we turn to the helical p-wave pairings ($A_{1u}$, $A_{2u}$, $B_{1u}$ and $B_{2u}$ irreps), which are analogues of the $^3$He B-phase~\cite{Vollhardt:90}. The four pairings individually preserve time-reversal symmetry, as each effectively consists of a pair of sub-Hilbert space characterized respectively by $k_x+ik_y$ and $k_x-ik_y$ pairings with identical gap amplitude. However, an appropriate complex mixture of them could tune the gap amplitude ratio between $k_x+ik_y$ and $k_x-ik_y$ away from unity (see Fig.~\ref{figSketch}); the resultant `asymmetric' state then has a net chirality and supports an intrinsic Kerr effect. Among all possible two-component order parameter combinations of the four helical pairings, we identify two that meet our requirement: $A_{1u}+iA_{2u}$ and $B_{1u}+iB_{2u}$. 

The idea is most transparent when presented in a single-orbital model. Take $A_{1u}+iA_{2u}$ as an example, in terms of the $\vec{d}$-vector, the pairing functions of $A_{1u}$ and $A_{2u}$ are given respectively by $k_x\hat{x}+k_y\hat{y}$ and  $k_x\hat{y}-k_y\hat{x}$. The full gap function then becomes
\begin{equation}
\vec{d}_{\bk} = \Delta_{A_{1u}}(k_x\hat{x}+k_y\hat{y}) + i \Delta_{A_{2u}} (k_x\hat{y}-k_y\hat{x}) \,,
\end{equation}
where $\Delta_{A_{1(2)u}}$ are real gap amplitudes. $\vec{d}_{\vec{k}}$ is non-unitary since $\vec{d}_{\vec{k}}^*\times \vec{d}_{\vec{k}} \ne 0$~\cite{Vollhardt:90}, implying unequal pairing gap amplitudes for the two spin species: $\Delta^{\uparrow\uparrow}(\vec{k})=-  (\Delta_{A_{1u}} - \Delta_{A_{2u}})(k_x - ik_y)$ for spin-up and $\Delta^{\downarrow\downarrow}(\vec{k})=(\Delta_{A_{1u}} + \Delta_{A_{2u}})(k_x + ik_y)$ for spin-down (Fig.~\ref{figSketch}). If $\Delta_{A_{1u}}= \Delta_{A_{2u}}$, the state resembles the $^3$He A$_1$-phase which has one spin component unpaired~\cite{Vollhardt:90}. More generally, both spin species are paired and they carry opposite orbital chiralities, i.e. opposite $L_z$. Such a state is odd-parity by nature~\cite{Nelson:04}, and it exhibits other salient features potentially consistent with the absence of spontaneous current signature~\cite{Kirtley:07,Hicks:10} and the substantial Knight shift drop~\cite{Pustogow:19,Ishida:20}, as we shall see later. Since the $B_{1u}+iB_{2u}$ state essentially shares the same general features, we will not present further analyses about it in the remaining of the paper.

At this point, it is worth pointing out that multiple microscopic calculations~\cite{Annett:06,Scaffidi:14,ZhangLD:18,Roising:19, Romer:19,Romer:20,WangZQ:20} have found that helical pairing states are favored in certain reasonable interaction parameter space. More interestingly, in some cases the splitting among various helical states could be rather small~\cite{WangZQ:20}, making it reasonable to consider accidentally degenerate pairing. Here, we make no attempt to delve into the pairing mechanism or how the pairing is distributed among different orbitals in \SRO. Instead, we will focus on the general phenomena that do not rely on microscopic details. Apart from some occasional general discussions, our analyses will be largely based on a concrete two-orbital model containing the Ru $d_{xz}$ and $d_{yz}$ orbitals. As far as the physical properties to be discussed are concerned, we expect the same qualitative features for more general multi-orbital models of \SRO. For simplicity, we assume the simple pairing functions for the various odd-parity (p-wave) superconducting channels as given in Table~\ref{table1}. 

\begin{table}[htbp]
    \centering
    \caption{Representative simple odd-parity pairing gap functions $(\vec{d}$-vectors) for the two-orbital model containing the Ru $d_{xz}$ and $d_{yz}$ orbitals. }
    \begin{tabular}{c  c  c}
        \hline
        ~~~~~~~~ irrep ~~~~~~~~& $~~~~~~~~\vec{d}_{xz}$ ~~~~~~~~& ~~~~~~~~$\vec{d}_{yz}$~~~~~~~~  \\
        \hline
        $A_{1u}$ & $\sin k_x \hat{x}$ & $\sin k_y \hat{y}$\\
        \hline
        $A_{2u}$ & $\sin k_x \hat{y}$ & $-\sin k_y \hat{x}$  \\
        \hline
        $B_{1u}$ & $\sin k_x \hat{x}$ & $-\sin k_y \hat{y}$   \\
        \hline
        $B_{2u}$ & $\sin k_x \hat{y}$ & $\sin k_y \hat{x}$ \\
        \hline
        $E_u$ &  $\sin k_x \hat{z}$ & $\sin k_y \hat{z}$ \\
        \hline
    \end{tabular}
    \label{table1}
\end{table}

\section{ac Anomalous Hall effect and Kerr rotation}
Since our proposed state is essentially composed of a pair of (spinless) chiral p-wave models with opposite chirality and inequivalent gap amplitudes, we expect it to support the anomalous Hall effect. Notably, while the intrinsic anomalous Hall effect is generally absent in single-orbital chiral superconducting models~\cite{Read:00,Taylor:12, Wysokinski:12,WangZQ:17}, \SRO~has an inherent multi-orbital character. 

Below we demonstrate intrinsic a.c.~anomalous Hall effect and Kerr rotation for the $A_{1u}+iA_{2u}$ state, using the same two-orbital model as in Ref.~\cite{Taylor:12}. 
Without qualitative impacts on our results, we neglect spin-orbital coupling (SOC) in this section for simplicity. 
Then the spin up and down blocks decouple, and their contributions to the two-dimensional Hall conductivity $\sigma_{\text{H}}$ can be computed separately. Here, we briefly describe the spin up block of the model. Written in the basis $(c_{d_{xz},\uparrow}(\vec{k}), c_{d_{yz},\uparrow}(\vec{k}))$, where $c_{d_{xz},\uparrow}$ and $c_{d_{yz},\uparrow}$
are annihilation operators for the $d_{xz}$ and $d_{yz}$ orbitals of $\mathrm{Sr_2RuO_4}$, 
the normal state Hamiltonian reads,
\begin{gather}
H_{\text{N}}(\vec{k})= 
\begin{pmatrix}
\xi_{xz}(\vec{k}) & \lambda(\vec{k}) \\
\lambda(\vec{k}) & \xi_{yz}(\vec{k})
\end{pmatrix},
\label{eq:HN}
\end{gather}
where $\xi_{xz}(\vec{k})=- 2 t \cos k_x -2\tilde{t}\cos(k_y)-\mu $, $\xi_{yz}(\vec{k})=- 2 \tilde{t} \cos k_x -2t\cos(k_y)-\mu$, $\lambda(\vec{k})=4 t^\prime \sin k_x \sin k_y$. Here $t$ labels the nearest neighbor hopping integral of the $d_{xz(yz)}$ orbital along the $x(y)$ direction, $\tilde{t}$ that of the $d_{xz(yz)}$ orbital along the $y(x)$ direction, $t^{\prime}$ denotes the hybridization of the two orbitals between next nearest neighboring sites. The corresponding BdG Hamiltonian follows as,
\begin{gather}
H_{\text{BdG}}^{\uparrow} (\vec{k})= 
\begin{pmatrix}
H_{\text{N}}(\vec{k})  &   \hat{\Delta}^{\uparrow\uparrow}(\vec{k})\\
[\hat{\Delta}^{\uparrow\uparrow}]^\dagger(\vec{k})  & - H_{\text{N}}^{T}(-\vec{k}) 
 \end{pmatrix},
\end{gather}
where, following Table \ref{table1},
\begin{gather}
\hat{\Delta}^{\uparrow \uparrow} (\vec{k})= 
 - (\Delta_{A_{1u}} -\Delta_{A_{2u}}) 
\begin{pmatrix}
 \sin k_x   &     0     \\
0       					&   -  i \sin k_y
\end{pmatrix}.
\end{gather}

\begin{figure}[htp]
\centering
\includegraphics[width=0.9\linewidth,trim={0 5mm 0 0}]{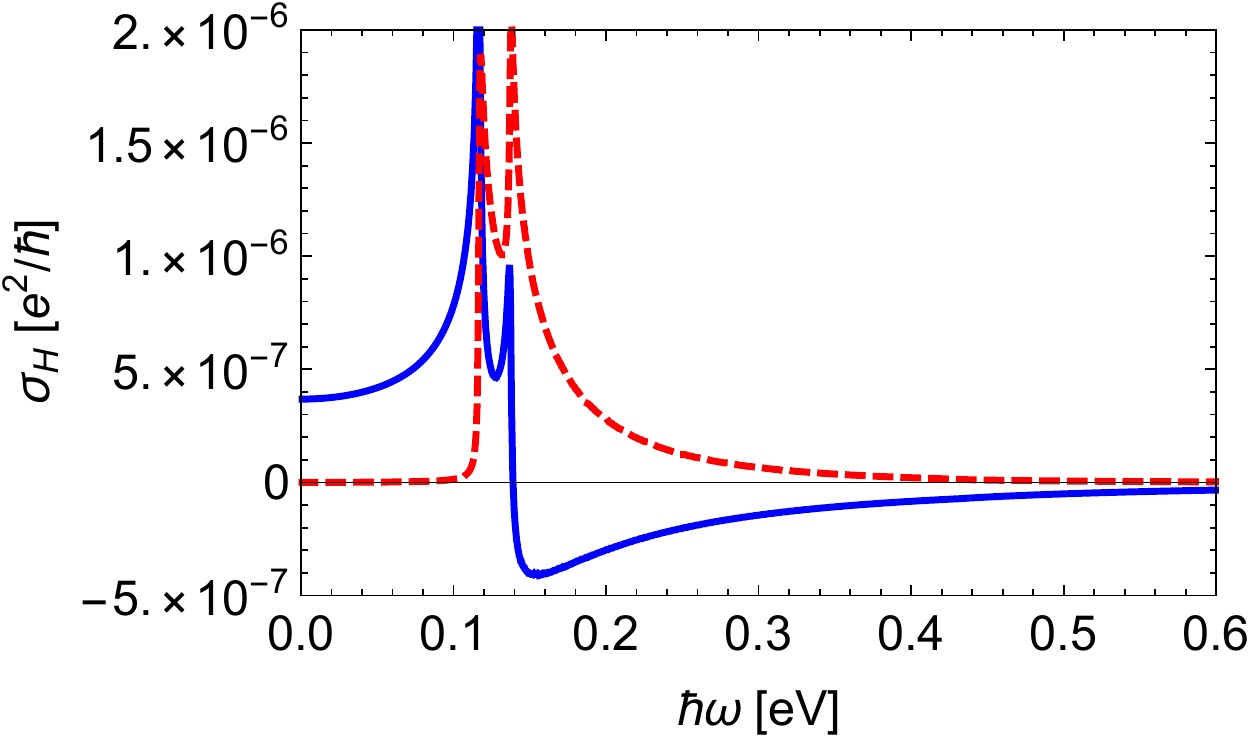}
\caption{(color online) Real (blue solid line) and imaginary (red dashed line) parts of the $\sigma_{\text{H}}(\omega)$ as a function of $\omega$ for $T=0$.
Band parameters are~\cite{Taylor:12}: $t=\mu= 0.4 \; \mathrm{eV}$, $t^\prime=0.05 t$ and $\tilde{t}=0.1 t$.
$\Delta_{A_{1u}}=2\Delta_{A_{2u}}=\frac{2}{3} \Delta_{\max}$, with $\Delta_{\max}=|\Delta_{A_{1u}}|+|\Delta_{A_{2u}}|=0.23 \; \mathrm{meV}$~\cite{Taylor:12}. 
We choose $|\Delta_{A_{1u}}|\ne |\Delta_{A_{2u}}|$ so that the gap magnitudes are nonzero for both spin up and down;
otherwise, one spin component would remain normal at zero temperature, which is not observed.
}
\label{fig: SigmaHTot}
\end{figure}

Using $H_{\text{BdG}}^{\uparrow}$ we calculate $\sigma_{\text{H}}^{\uparrow}(\omega)$ from the standard Kubo formula (without vertex corrections). Details of the calculation can be found in the Supplementary Materials~\cite{SM}. The same calculation is done for $\sigma_{\text{H}}^{\downarrow}$ and the numerical result of the total $\sigma_{\text{H}}$ for temperature $T=0$ is presented in Fig.~\ref{fig: SigmaHTot}. Not surprisingly, $\sigma_{\text{H}}$ is nonzero; both its real and imaginary parts have similar frequency dependence to those obtained for a chiral p-wave pairing~\cite{Taylor:12}. The overall magnitude of $\sigma_{\text{H}}$ is reduced by a factor of 2
compared to that in Ref.~\cite{Taylor:12}, because in the current scenario $\sigma_{\text{H}}^{\uparrow}$  and $\sigma_{\text{H}}^{\downarrow}$ partially cancel each other due to
the opposite chirality of $\hat{\Delta}^{\uparrow \uparrow}$ and $\hat{\Delta}^{\downarrow \downarrow}$, which dictates the signs of $\sigma_{\text{H}}^\uparrow$ and $\sigma_{\text{H}}^\downarrow$, respectively. On the other hand, such a cancellation is absent in the chiral p-wave pairing case. 

We note that $\mathrm{Im}[\sigma_{\text{H}}(\omega)]$ becomes nonzero at $\omega \gtrsim 0.4 \, \mathrm{eV}$ for the chosen band parameters.
This frequency is determined by the minimum energy cost to create a pair of Bogoliubov quasiparticles from different bands, which is
not dictated by $2| \Delta^{\uparrow \uparrow}|$ or $2|\Delta^{\downarrow \downarrow}|$ but by the hopping parameter $t^\prime$~\cite{Taylor:12}. 
Therefore, even though $| \Delta^{\uparrow \uparrow}|$ and $|\Delta^{\downarrow \downarrow}|$ can be quite different, the onset frequencies of $\mathrm{Im}[\sigma_{\text{H}}^{\uparrow}]$ and $\mathrm{Im}[\sigma_{\text{H}}^{\downarrow}]$ are actually almost identical. 

From the calculated $\sigma_{\text{H}}$ one can estimate the Kerr rotation angle using
\begin{gather}
\theta_{\text{K}}(\omega) = \frac{4\pi}{\omega d} \; \mathrm{Im}[ \frac{ \sigma_{\text{H}}(\omega) } { n(n^2-1)}],
\end{gather}
where $n=n(\omega)$ is the complex index of refraction and $d$ is the interlayer spacing of Sr$_2$RuO$_4$
along the $c$-axis.  $n(\omega)$ can be estimated from the longitudinal optical conductivity which is
modeled by a Drude model as in Ref.~\cite{Taylor:12}. For details see Ref.~\onlinecite{SM}. 
The estimated $\theta_{\text{K}}(\hbar \omega=0.8~\text{eV}) \approx 20 \; \mathrm{nrad}$~\cite{SM}, which may potentially account for the experimental value of
$65 \; \mathrm{nrad}$ observed by J. Xia \textit{et al}~\cite{Xia:06}, given the uncertainty in our estimate of $n(\omega)$ as well as
the neglecting of the $\gamma$ band in the current calculation. We note that $\sigma_{\text{H}}$, and therefore the estimated Kerr angle $\theta_{\text{K}}$,
does depend on the actual ratio between $|\Delta_{A_{1u}}|$ and $|\Delta_{A_{2u}}|$. However, the dependence is quite weak
as long as the two are comparable~\cite{SM}.

\section{Spontaneous current}
A chiral p-wave state shall generate finite spontaneous surface current~\cite{Matsumoto:99,Furusaki:01}. Following the original proposal of chiral p-wave pairing in \SRO, attempts to detect signatures of such a current have had little success, suggesting that the spontaneous current is either simply absent or too tiny to be resolved in actual measurements~\cite{Kirtley:07,Hicks:10}. 

The $A_{1u} + iA_{2u}$ state has just the right appeal for explaining this, which is already obvious in the single-orbital case. Here, the two spin species each generates an edge current, and these currents flow in opposite direction due to the opposite chirality. Moreover, the integrated current in fact vanishes in the BCS limit. However, thanks to the distinct decay length scales of the two contributions, the current distribution does not exactly cancel at each spatial point. 

\begin{figure}
\includegraphics[width=8.cm, trim={0 8mm 0 0}]{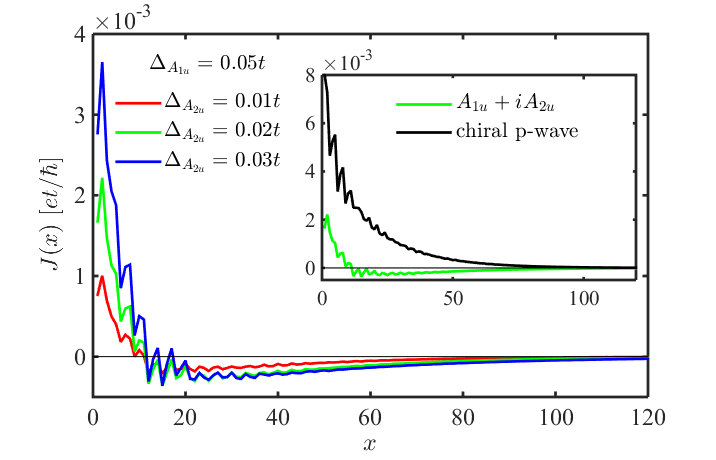}
\caption{(color online) Distribution of zero-temperature spontaneous edge current in our two-orbital model with a mixed helical p-wave $A_{1u}+iA_{2u}$ pairing. The $A_{1u}$ component is held fixed at $\Delta_{A_{1u}}=0.05t$. Inset: a comparison of the spontaneous current of the mixed helical and chiral p-wave states with similar gap amplitudes. These calculations follow those in Refs.~\cite{Imai:12,Imai:13,Lederer:14}, and the lattice constant is defined to be unity.}
\label{figCurr}
\end{figure}

Going beyond single-orbital model, the integrated current may not vanish due to SOC and other band structure effects. However, the strong suppression persists, as is demonstrated in Fig.~\ref{figCurr} for our two-orbital model. Neglecting the fast Friedel oscillations, the spontaneous current generally contains two canceling contributions characterized by different decay length scales. For the calculations shown in the figure, the net current in the mixed helical state is typically more than 20 times smaller than that in a chiral p-wave state with similar gap magnitudes (see the inset of Fig.~\ref{figCurr}). 

It should be stressed that such suppression is inherent, as it is achieved without appealing to surface disorder, anisotropic band and gap structure --- factors typically having the potential to further reduce the spontaneous current, as is known for the case of chiral p-wave~\cite{Ashby:09,Imai:12,Imai:13,Bouhon:14,Lederer:14,Huang:15,Scaffidi:15}. In fact, this behavior is also expected from the Ginzburg-Landau theory, at the lowest order of which the spontaneous current is related to cross-gradient bilinears of the involving order parameters, such as $\partial_x \Delta_{A_{1u}}^\ast \partial_y \Delta_{A_{2u}}$~\cite{Sigrist:91,Huang:15}. In the present case, this lowest order contribution vanishes by symmetry~\cite{SM}, and higher order terms such as $\partial_x^3 \Delta_{A_{1u}}^\ast \partial_y \Delta_{A_{2u}}$ are needed to account for the small local current distribution. By the same token, similar physics applies to the chiral $d_{x^2-y^2}+id_{xy}$ pairing~\cite{WangX:18}. In passing, we note that for the $d_{x^2-y^2}+ig_{xy(x^2-y^2)}$ pairing~\cite{Kivelson:20} the spontaneous current is in general finite at the (100) surfaces and vanishes at the (110) surface, analogous to the situation in $s+id_{xy}$ pairing~\cite{Furusaki:01}. 

Finally, the suppression in the $A_{1u}+ i A_{2u}$ phase is also independent of the edge orientation, and the same phenomenology takes place at the domain walls separating regions of time-reversed pairings.

\section{Spin susceptibility and Knight Shift}
Recent revised NMR measurements under external in-plane magnetic field reveal a substantial Knight shift drop below $T_c$~\cite{Pustogow:19,Ishida:20}. This implies a drop of spin susceptibility, $\chi_\text{spin}$, which was subsequently confirmed by a polarized neutron scattering study~\cite{Petsch:20}. A later NMR study further places an upper bound of $10\%$ of the normal state susceptibility at the lowest measured temperatures~\cite{Chronister:20}. These reports can be most straightforwardly explained in terms of spin-singlet pairing. However, as the observed Knight shift contains contributions from both spin and orbital degrees of freedom, the exact fraction of the spin susceptibility drop in the zero field limit may deserve a more careful inquiry~\cite{Leggett:20}. As we argue below, our proposed states may still have the potential to reconcile with the susceptibility drop. 

\begin{figure}
\includegraphics[width=8.cm,trim={0 8mm 0 0}]{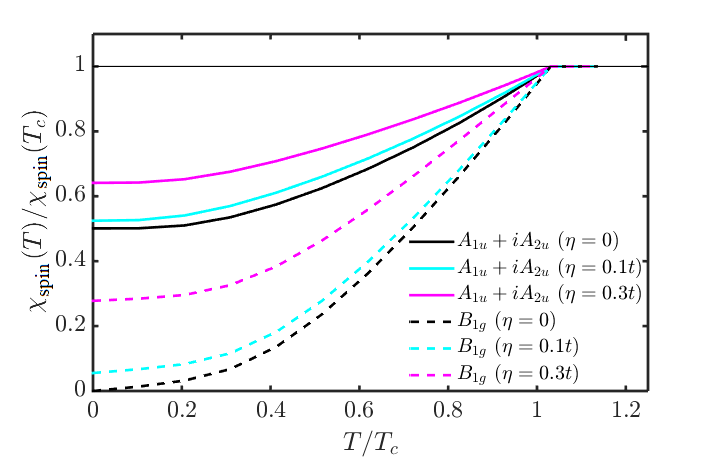}
\caption{(color online) Temperature dependence of the spin susceptibility in the $A_{1u}+iA_{2u}$ (solid curves) and $B_{1g}$ (dashed curves) superconducting states, in the presence of a small Zeeman field in the $x$-direction. Here, $\eta$ denotes the strength of SOC.}
\label{figSus}
\end{figure}

Owing to the in-plane $\vec{d}$-vector orientation, the mixed helical states naturally exhibit a reduced $\chi_\text{spin}$ across $T_c$ under the in-plane field. In fact, in the absence of SOC, $\chi_\text{spin}$ at $T=0$ equals half of its normal state value. This is demonstrated in Fig.~\ref{figSus}, which shows the temperature-dependent susceptibility in the presence of a weak Zeeman field directed along the $x$-direction. For comparison, we also show results for a spin-singlet d-wave state (a pairing in the $B_{1g}$ irrep). In the calculations, we have assumed that the two helical pairings onset at the same critical temperature, and that their temperature dependence follows the standard BCS mean-field behavior. 

The SOC may facilitate extra spin-flipping processes that could have otherwise been suppressed by Cooper pairing, thereby alleviating the drop of $\chi_\text{spin}$. This is exemplified by the noticeable increase of $\chi_\text{spin}(T=0)$ with enhanced SOC (Fig.~\ref{figSus}). Notably, even a pure spin-singlet pairing may develop a residual $\chi_\text{spin}(T=0)$~\cite{Roising:19, Romer:19}. For the two-orbital model in question, this can be explained in the following terms. Here, SOC takes the form of $\eta \hat{L}_z \hat{s}_z$ where $\hat{L}_z$ operates on the orbital degrees of freedom and $\hat{s}_z$ is the spin Pauli matrix. 
To be specific, $\hat{L}_z=- i \epsilon_{i j}$~\cite{WangZQ:20} with $\epsilon_{ij}$ the fully antisymmetric tensor and $\{i,j\}=\{1,2\}$ representing the $\{d_{xz}, d_{yz}\}$ orbitals in Eq.~\eqref{eq:HN}. The SOC term flips the in-plane spin components but not the ones along $z$. As a consequence, upon turning on a finite $\eta$, the $z$-component of the spin susceptibility $\chi^{zz}_\text{spin}$ remains fully suppressed at $T=0$ (not shown in Fig.~\ref{figSus}), whereas the $x$-component, $\chi^{xx}_\text{spin}$, acquires a residual value. We expect such an enhancement of $\chi_\text{spin}(T=0)$ due to SOC to persist in a more general multi-orbital model of \SRO~\cite{Huang:19,Ramires:19,Kaba:19}. Finally, it is worth remarking that, the natural mixing of (real)-spin triplet and singlet pairings in the presence of SOC~\cite{Puetter:12,Huang:19}, which is not considered here, may also introduce corrections to the spin susceptibility. 

Note that our calculations were done in the non-interacting limit. Correlations renormalize the normal and superconducting state $\chi_\text{spin}$ in different manners, reducing the ratio $\chi_{\text{spin}}(T=0)/\chi_{\text{spin}}(T=T_c)$ from its non-interacting value~\cite{Leggett:65}. Based on an estimate of the Wilson ratio of 2~\cite{Steffens:19,Ishida:20} and using the data for $\eta=0.1t$ in Fig.~\ref{figSus}, we obtain $\chi_{\text{spin}}(T=0)/\chi_{\text{spin}}(T=T_c) \approx 35\%$ for the $A_{1u}+iA_{2u}$ state. This is a substantial overall suppression, although, at face value, still noticeably higher than the upper bound of $10\%$ suggested by the most recent NMR measurement~\cite{Chronister:20}. However, our crude approximation could be insufficient for a quantitative comparison, and a calculation based on a more realistic correlated multiorbital model may be needed. Furthermore, the potential complications in disentangling the orbital and spin contributions to the Knight shift~\cite{Pavarini:07,Leggett:20} could affect the experimental interpretation. 

\section{Strain and ultrasound}
Uniaxial strain and ultrasound measurements reveal crucial information about the symmetry of the superconducting order~\cite{Hicks:14,Ghosh:20,Benhabib:20}. Most informative among the existing experiments are the ones that probe the coupling between the order parameters and the $B_{1g}$ ($\epsilon_{x^2-y^2}$) and $B_{2g}$ ($\epsilon_{xy}$) strains. For our mixed helical state, the leading order coupling has the following form in the Ginzburg-Landau free energy, 
\begin{equation}
f_\text{coupling}  = \sum_i \epsilon_i^2 (\alpha_i |\Delta_{A_{1u}}|^2 + \beta_i |\Delta_{A_{2u}}|^2)
\label{eq: fcoupling}
\end{equation}
where $i=x^2-y^2, xy$ for the $B_{1g}$ and $B_{2g}$ strains, respectively, and $\alpha_i (\beta_i)$ are coupling constants. No other order parameter bilinears are present. This free energy has two implications. On the one hand, the mean-field critical temperatures of the two components shall both follow a quadratic dependence on the strain. This seems to agree with the observed behavior in several uniaxial strain measurements~\cite{Steppke:17,Watson:18}. We do note that there is no solid evidence of the expected split transitions in thermodynamic measurements~\cite{Yonezawa:14,LiYS:19}, although this could be explained away by arguments unrelated to symmetry~\cite{Scaffidi:20}. On the other hand, the absence of a linear coupling to $\epsilon_i$ in Eq.~\eqref{eq: fcoupling} disallows any discontinuity in the shear elastic moduli $(c_{11}-c_{12})/2$ or $c_{66}$. This stands in contrast with recent reports, where a jump in the latter has been observed~\cite{Ghosh:20,Benhabib:20}. Interestingly, a linear coupling to the $B_{2g}$ strain (associated with $c_{66}$) is possible for mixed $A_{1u}$ and $B_{2u}$, as well as mixed $A_{2u}$ and $B_{1u}$ states~\cite{WangZQ:20}. This suggests a way to reconcile our proposed order with the ultrasound experiments: the breaking of crystalline rotational symmetry around lattice dislocations can allow the $B_{1u}$ and/or $B_{2u}$ order parameters to locally condense against the backdrop of a bulk $A_{1u}+iA_{2u}$ order, leading to a local mixture of
$A_{1u}$ with $B_{2u}$ (and/or $A_{2u}$ with $B_{1u}$) that couples linearly to the $B_{2g}$ strain.

\section{Concluding remarks}
Under the assumption that the observed polar Kerr effect is a genuine response of superconducting \SRO~in the clean limit, we proposed that the mixed helical p-wave states, $A_{1u}+iA_{2u}$ and $B_{1u}+iB_{2u}$, represent faithful alternative candidates to chiral superconducting states. We also discussed how these states may be compatible with several other key measurements. 
Interestingly, a recent study~\cite{Gupta:20} shows that a helical state may also explain the Pauli limiting behavior of the in-plane $H_{c2}$ and the first order phase transition observed in the $H-T$ phase diagram at low $T$ and at $H$ near $H_{c2}$~\cite{Deguchi:02,Yonezawa:13}. Nevertheless, a full reconciliation with other important observations may require further examination. For example, the presence of gapless excitations~\cite{Nishizaki:00,Deguchi:04,Firmo:13,Hassinger:17,Dodaro:18} would place strong constraint on the detailed forms of the p-wave pairings. To this end, we take note of the multitude of possibilities made available by the multiorbital nature of this material~\cite{Agterberg:97,Zhitomirsky:01,Ramires:16,Ramires:19,Huang:19,Kaba:19,WangWS:19,Gingras:19,LiYu:19,Suh:19,ChenW:20,Lindquist:19}.

Due to their non-unitary nature, our proposed states are also expected to exhibit finite spin polarization in the bulk, which scales as $(\Delta/E_F)^2$. The resultant spontaneous bulk magnetization may be probed in SQUID measurements. One such attempt was said to be impeded by the Meissner effect~\cite{Sumiyama:13}, while the existing scanning SQUID~\cite{Kirtley:07,Hicks:10} techniques may need improved precision to resolve the resultant weak magnetization on the sample surface. 

As a final remark, the $B_{1g}+iB_{2g}$ state (chiral $d_{x^2-y^2} + id_{xy}$ pairing) also carries the essential features needed to explain the various observations discussed in this work. Alternatively, in a spirit similar to that proposed in Ref.~\cite{Willa:20}, one can consider that one component of the $B_{1g}+iB_{2g}$ state is favored in the clean limit, while the other component condenses around lattice dislocations.

\textit{Note added}. After posting our manuscript on arXiv, we were informed of another work~\cite{Hu:21}, which discusses the spontaneous magnetization of the $A_1+iA_2$ and $B_1+iB_2$ superconducting states in a non-centrosymmetric system with $C_{4v}$ point group symmetry. 

\section{Acknowledgements} We acknowledge helpful discussions with Weiqiang Chen, Catherine Kallin, Yongkang Luo, Aline Ramires, Thomas Scaffidi and Fuchun Zhang. WH is supported by NSFC under grant No.~11904155, the Guangdong Provincial Key Laboratory under Grant No.~2019B121203002, and a Shenzhen Science and Technology Program (KQTD20200820113010023).  ZW acknowledges financial support from the James Franck Institute at the University of Chicago. Computing resources are provided by the Center for Computational Science and Engineering at Southern University of Science and Technology.

\end{document}




\title{Supplemental Material for ``Possibility of mixed helical p-wave pairings in Sr$_2$RuO$_4$"}
\author{Wen Huang}
\affiliation{Shenzhen Institute for Quantum Science and Engineering \& Guangdong Provincial Key Laboratory of Quantum Science and Engineering, Southern University of Science and Technology, Shenzhen 518055, Guangdong, China}
\author{Zhiqiang Wang}
\affiliation{James Franck Institute, University of Chicago, Chicago, Illinois 60637, USA}
\date{\today}

\maketitle

\section{Calculation of the anomalous Hall conductivity}
In the absence of SOC, the two spin blocks decouple. We can calculate $\sigma_{\text{H}}^\uparrow$ and $\sigma_{\text{H}}^\downarrow$ separately. 

\subsection{Spin up block}
\label{sec:spinup}
The model we consider consists of the two quasi-1d orbitals, $d_{xz}$ and $d_{yz}$, of \SRO.
Written in the basis $\{c_{d_{xz}}(\vec{k}), c_{d_{yz}}(\vec{k})\}$ the spin up block of the normal state Hamiltonian reads
\begin{gather}
\hat{H}_{\text{N}}(\vec{k})
\equiv \begin{pmatrix}
\xi_{xz}(\vec{k})   & \lambda(\vec{k})   \\ 
\lambda(\vec{k})   & \xi_{yz}(\vec{k})  
\end{pmatrix}
= 
\begin{pmatrix}
- 2 t \cos k_x -2 \tilde{t} \cos k_y  -\mu & 4 t^\prime \sin k_x \sin k_y \\
4 t^\prime \sin k_x \sin k_y & -2 t \cos k_y - 2 \tilde{t} \cos k_x -\mu 
\end{pmatrix}.
\end{gather}
The corresponding BdG Hamiltonian is 
\begin{gather}
\hat{H}_{\text{BdG}}^{\uparrow} (\vec{k})
= 
\begin{pmatrix}
\hat{H}_{\text{N}}(\vec{k})  &   \hat{\Delta}^{\uparrow\uparrow}(\vec{k})\\
[ \hat{\Delta}^{\uparrow\uparrow}]^\dagger(\vec{k})  & - [\hat{H}_{\text{N}}]^{T}(-\vec{k}) 
 \end{pmatrix}
 \label{eq: HBdG}
\end{gather}
with
\begin{gather}
\hat{\Delta}^{\uparrow \uparrow} (\vec{k})
=
- (\Delta_{A_{1u}} -\Delta_{A_{2u}}) 
\begin{pmatrix}
 \sin k_x   &     0     \\
0       					&   - i \sin k_y
\end{pmatrix},
\end{gather}
which is effectively a (spinless) chiral $p_x - i p_y$ order parameter. 
For simplicity we have considered only intra-orbital pairing while neglected any inter-orbital one. 
The $d_{xz}$ ($d_{yz}$) intraorbital component $\Delta^{\uparrow\uparrow}_{ 11}$ ($\Delta^{\uparrow\uparrow}_{22}$) transforms like
$k_x$ ($k_y$) under the spatial $D_{\text{4h}}$ group.
Hence, $\Delta^{\uparrow\uparrow}_{ 11}(\vec{k}) = \big[\vec{d}_{\vec{k}}^{(x)}\cdot \boldsymbol{\sigma} i \sigma_2 \big]_{\uparrow \uparrow} =-(\Delta_{A_{1u}} -\Delta_{A_{2u}})  \sin k_x$,
where $\vec{d}_{\vec{k}}^{(x)}$ denotes the $k_x$-like part of $\vec{d}_{\vec{k}} $. Note that $\vec{d}_{\vec{k}} = \Delta_{A_{1u}} (\hat{x} \sin k_x + \hat{y} \sin k_y) + \Delta_{A_{2u}}(\hat{x} \sin k_y -\hat{y} \sin k_x)$.
Similarly,  $\Delta^{\uparrow\uparrow}_{ 22}(\vec{k}) = \big[\vec{d}_{\vec{k}}^{(y)}\cdot \boldsymbol{\sigma} i \sigma_2 \big]_{\uparrow \uparrow}$. 
From $H_{\text{BdG}}$ we can define the electric current velocity operator 
\begin{gather}
\hat{\vec{v}}(\vec{k})
=
\begin{pmatrix}
\nabla_{\vec{k}} \xi_{xz}(\vec{k}) & \nabla _{\vec{k}} \lambda(\vec{k}) \\
\nabla_{\vec{k}} \lambda(\vec{k}) & \nabla_{\vec{k}} \xi_{yz}(\vec{k})
\end{pmatrix}
\otimes 1_{2\times2},
\end{gather}
where $1_{2\times2}$ is the identity matrix in the Nambu particle-hole space. 

We calculate $\sigma_{\text{H}}^{\uparrow} $ using the standard one-loop Kubo formula
\begin{align}
\sigma_{\text{H}}^{\uparrow} (\omega) & =  \frac{i}{\omega} \frac{\pi_{xy}(\vec{q}=0,\omega+i 0^+) - \pi_{yx}(\vec{q}=0,\omega+ i 0^+)}{2},
\end{align}
where $\pi_{xy}(\vec{q}, i \nu_m)$ is the electric current-current density correlator, given by
\begin{align}
\pi_{xy}(0, i\nu_m )  & = e^2 T \sum_n \sum_{\vec{k}} \mathrm{Tr}[\hat{v}_x(\vec{k}) \hat{G}(\vec{k},i\omega_n) \hat{v}_y(\vec{k}) \hat{G}(\vec{k},i\omega_n + i \nu_m) ].
\label{eq: pixy}
\end{align}
In this equation $\hat{G}(\vec{k},i\omega_n) \equiv [ i\omega_n - \hat{H}_{\text{BdG}}^{\uparrow}(\vec{k})]^{-1}$ is the Green's function. 
$\omega_n = (2n +1)\pi T$ and $\nu_m=2 m \pi T$ are Fermionic and Bosonic Matsubara frequencies.
The trace $\mathrm{Tr}$ is with respect to both the Nambu particle-hole and orbital spaces. 
$e$ is the charge of an electron. 
Using Eq.~\eqref{eq: HBdG} in the expression of $\pi_{xy}(0,i \nu_m)$ in Eq.~\eqref{eq: pixy}, completing the Matsubara sum over $\omega_n$,
and then making the analytical continuation, $i \nu_m \rightarrow \omega + i \delta$, we obtain
\begin{gather}
\frac{\sigma_{\text{H}}^{\uparrow}(\omega) }{e^2/\hbar}=  \sum_{\vec{k}}
 \frac{ \mathcal{F}^{\uparrow}_{\vec{k}} }{E_{+}^\uparrow(\vec{k})  E_{-}^\uparrow(\vec{k})  [E_{+}^\uparrow(\vec{k})  + E_{-}^\uparrow(\vec{k})]  [(E_{+}^\uparrow(\vec{k})  + E_{-}^\uparrow(\vec{k}) )^2 -( \omega+i\delta)^2]},
\end{gather}
where
\begin{align}
\mathcal{F}^{\uparrow}_{\vec{k}} 
& = 32 \;  (t-\tilde{t}) (t^\prime)^2 (\Delta_1 - \Delta_2)^2  (\sin^2 k_x \cos k_y + \sin^2 k_y \cos k_x) \sin^2 k_x \sin^2 k_y.
\label{eq: FkUp}
\end{align}
$E_{\pm}^\uparrow(\vec{k})$ are eigenvalues of $\hat{H}_{\text{BdG}}^{\uparrow}(\vec{k})$. 
\begin{subequations}
\label{eq: EpmUp}
\begin{align}
E_{\pm}^\uparrow & = \sqrt{ \frac{-\alpha \pm \sqrt{\alpha^2 - 4 \beta}}{2}}, \\
\alpha & = - (\xi_{xz}^2 + |\Delta^{\uparrow\uparrow}_{11}|^2 + \xi_{yz}^2 + |\Delta^{\uparrow \uparrow}_{22}|^2 + 2 \lambda^2), \\
\beta  & = (\xi_{xz}^2 +  |\Delta^{\uparrow\uparrow}_{ 11}|^2)  (\xi_{yz}^2 +  |\Delta^{\uparrow\uparrow}_{ 22}|^2) + \lambda^4 + 
\lambda^2 ( [\Delta^{\uparrow \uparrow}_{11}]^* \Delta^{\uparrow \uparrow}_{ 22} + \Delta^{\uparrow\uparrow}_{ 11} [\Delta^{\uparrow\uparrow}_{ 22}]^*)
-2 \lambda^2 \xi_{xz} \xi_{yz}.
\end{align}
\end{subequations}
For brevity we have suppressed the $\vec{k}$ dependence in these equations. 

Written out explicitly, 
\begin{align}
\mathrm{Re}[   \sigma_{\text{H}}^\uparrow ] (\omega) & =  \frac{e^2}{\hbar} \sum_{\vec{k}} \frac{ \mathcal{F}^{\uparrow}_{\vec{k}}}{E_{ +}^\uparrow E_{-}^\uparrow (E_{+}^\uparrow  + E_{- }^\uparrow  )  [(E_{+}^\uparrow  + E_{-}^\uparrow )^2 - \omega^2]} , \\
\mathrm{Im}[   \sigma_{\text{H}}^{\uparrow}   ] (\omega) & =  \frac{e^2}{\hbar}  \sum_{\vec{k}}  \frac{ \mathcal{F}^{\uparrow}_{\vec{k}}}{E_{+}^\uparrow E_{-}^\uparrow }
 \frac{\pi }{2 \omega^2}  \bigg[ \delta( \omega - (E_{+}^\uparrow  + E_{-}^\uparrow )) - \delta(\omega + (E_{+ }^\uparrow  + E_{-}^\uparrow  ) ) \bigg].
 \label{eq: ImSigmaHUp}
\end{align}

\subsection{Spin down block}
\begin{figure}[h]
\centering
\includegraphics[width=0.45\linewidth]{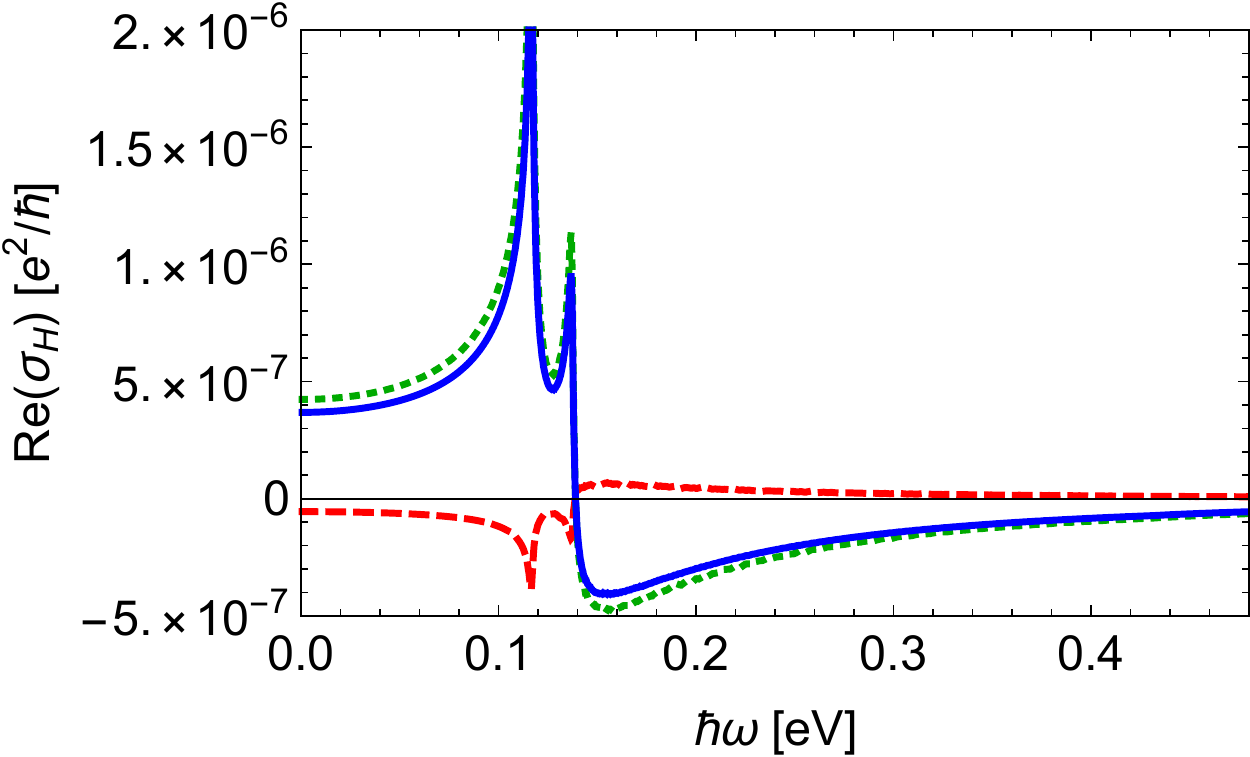} 
\hspace{3mm}
\includegraphics[width=0.45\linewidth]{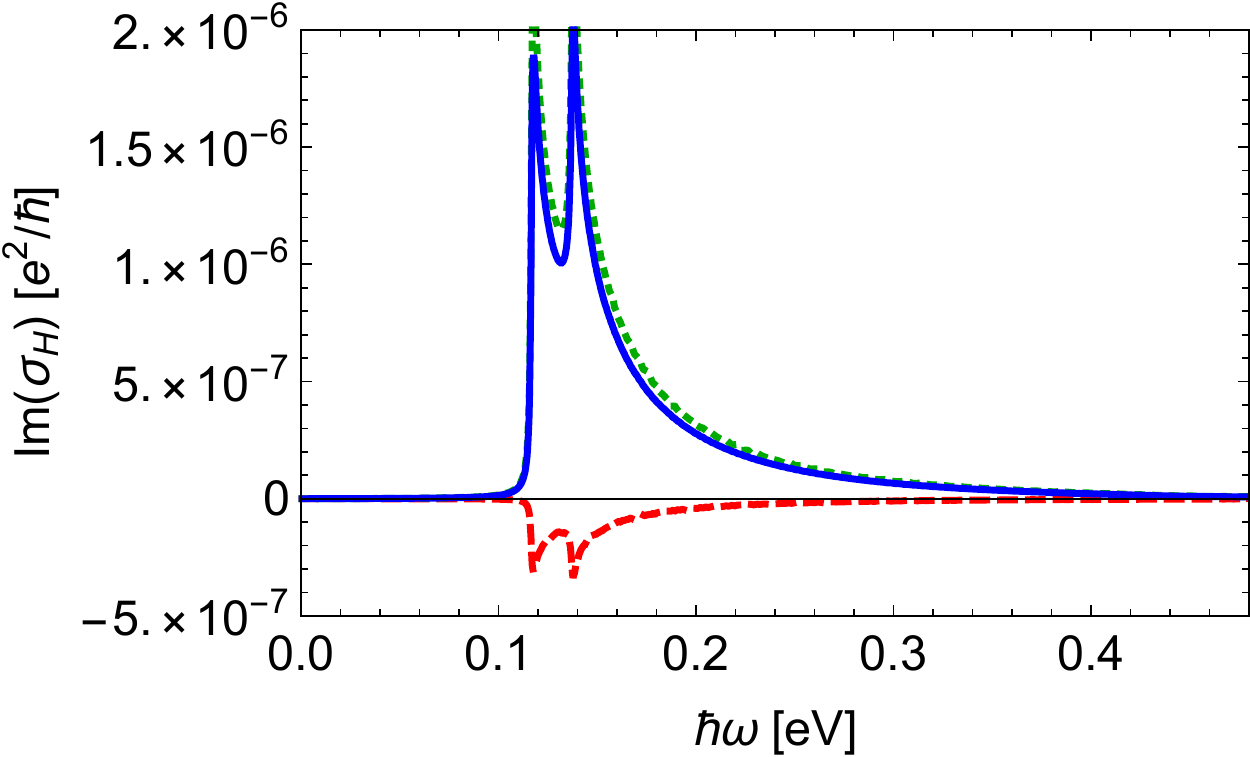}
\caption{ Spin $\uparrow$ (red dashed line), $\downarrow$ (dark green dotted line), and the total (blue solid line)
contributions to the $T=0$ Hall conductivity.
Left: $\mathrm{Re}[\sigma_{\text{H}}]$;
Right: $\mathrm{Im}[\sigma_{\text{H}}]$.
Band parameters used are~\cite{Taylor:12}: $t=\mu = 0.4 \, \mathrm{eV}$, $\tilde{t}=0.1 t$, and $t^\prime=0.05 t$.
$\Delta_{A_{1u}}=2\Delta_{A_{2u}}=\frac{2}{3} \Delta_{\max}$ with $\Delta_{\max}=0.23 \,\mathrm{meV}$~\cite{Taylor:12}. 
}
\label{fig: SigmaH}
\end{figure}

The derivation of $\sigma_{\text{H}}^{\downarrow}$ is almost identical. The only difference is that in the definition of the BdG Hamiltonian,
$\hat{\Delta}^{\uparrow\uparrow}$ is now replaced by 
\begin{gather}
\hat{\Delta}^{\downarrow \downarrow} (\vec{k})
=
(\Delta_{A_{1u}} + \Delta_{A_{2u}}) 
\begin{pmatrix}
\sin k_x   &     0     \\
0       					&   i \sin k_y
\end{pmatrix}.
\end{gather}
Correspondingly, 
\begin{gather}
\frac{\sigma_{\text{H}}^{\downarrow}(\omega) }{e^2/\hbar}=  \sum_{\vec{k}}
 \frac{\mathcal{F}^{\downarrow}_{\vec{k}}  }{E_{+}^\downarrow E_{-}^\downarrow (E_{+}^\downarrow  + E_{-}^\downarrow  )  [(E_{+}^\downarrow  + E_{-}^\downarrow )^2 -( \omega+i\delta)^2]}
\end{gather}
with 
\begin{align}
\mathcal{F}^{\downarrow}_{\vec{k}}  & = - 32  (t - \tilde{t}) (t^\prime)^2 (\Delta_1 + \Delta_2)^2  (\sin^2 k_x \cos k_y + \sin^2 k_y \cos k_x) \sin^2 k_x \sin^2 k_y. 
\end{align}
$E_{\pm}^\downarrow$ differs from $E_{\pm}^\uparrow$ in Eq.~\eqref{eq: EpmUp} by replacing $\{\Delta^{\uparrow \uparrow}_{ 11},\Delta^{\uparrow \uparrow}_{22}\}$
with $\{\Delta^{\downarrow \downarrow}_{11},\Delta^{\downarrow \downarrow}_{ 22}\}$. 
Notice that although the gap magnitudes can be quite different for spin up and down, $\min\{E_+^\uparrow(\vec{k}) + E_-^\uparrow(-\vec{k}) \}$, which determines the onset frequency
of $\mathrm{Im}[\sigma_{\text{H}}^\uparrow](\omega)$ (see Eq.~\eqref{eq: ImSigmaHUp}), is actually almost the same as $\min\{E_+^\downarrow(\vec{k}) + E_-^\downarrow(-\vec{k}) \}$, 
because both of them are governed by the orbital-hybridization parameter $t^\prime$~\cite{Taylor:12}, instead of the pairing gap magnitudes. 
This explains why, in Fig.~\ref{fig: SigmaH}, $\mathrm{Im}[\sigma_{\text{H}}^\uparrow](\omega)$ and $\mathrm{Im}[\sigma_{\text{H}}^\downarrow](\omega)$ become
nonzero essentially at the same frequency. 

\begin{figure}[h]
\centering
\includegraphics[width=0.5\linewidth]{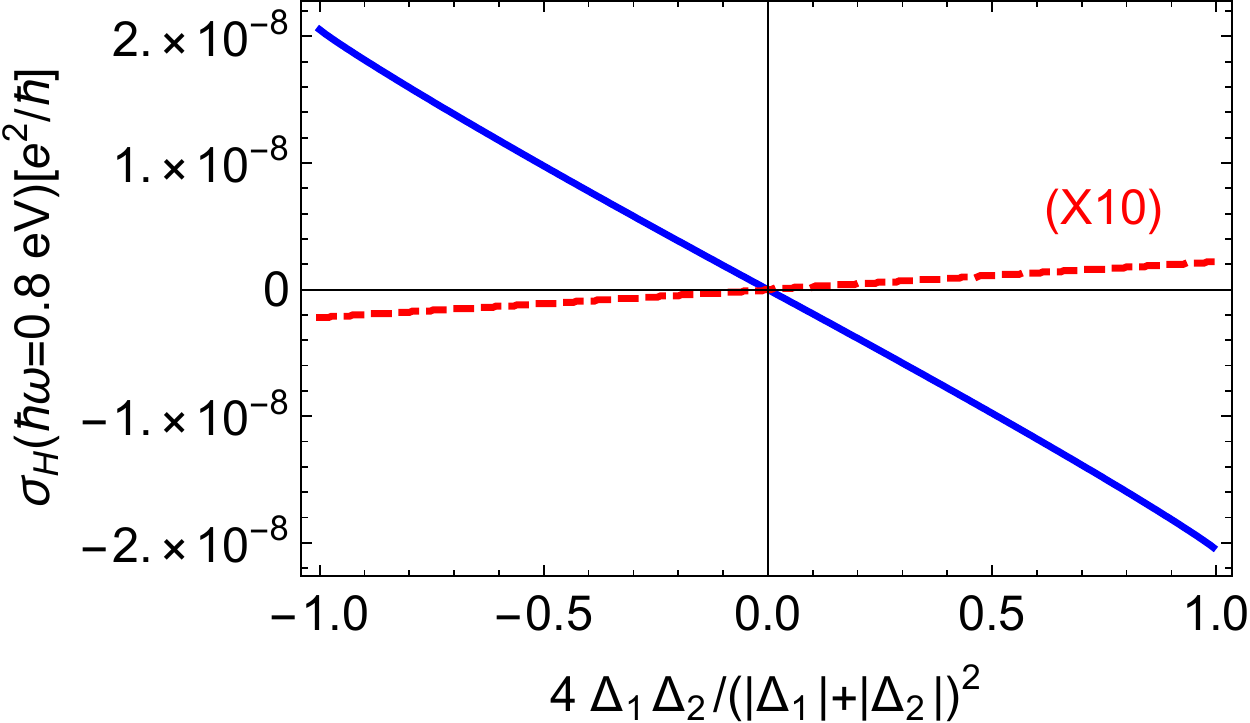}
\caption{ Real (blue solid line) and imaginary (red dashed line) parts of the zero temperature $\sigma_{\text{H}}(\omega)$
as a function of $4\Delta_{A_{1u}} \Delta_{A_{2u}}/(|\Delta_{A_{1u}}|+|\Delta_{A_{2u}}|)^2$. $\hbar \omega=0.8 \, \mathrm{eV}$. $\Delta_{\max}= |\Delta_{A_{1u}}|+|\Delta_{A_{2u}}| = 0.23 \,  \mathrm{meV}$ is kept
a constant in this calculation. The plot shows that $\sigma_{\text{H}}$ is roughly linear in $ 4\Delta_{A_{1u}} \Delta_{A_{2u}}/(|\Delta_{A_{1u}}|+|\Delta_{A_{2u}}|)^2$. 
From the plot we see that $\sigma_{\text{H}}$ is odd in the relative sign between $\Delta_{A_{1u}}$ and $ \Delta_{A_{2u}}$, but even under the interchange $|\Delta_{A_{1u}}| \leftrightarrow |\Delta_{A_{2u}}|$.
}
\label{fig: ImSigmaH0d8eV}
\end{figure}

Comparing $\mathcal{F}_{\vec{k}}^\downarrow$ to $\mathcal{F}_{\vec{k}}^\uparrow$ in Eq.~\eqref{eq: FkUp} we see that they carry opposite signs, leading to a partial cancellation between $\sigma_{\text{H}}^{\downarrow}$
and $\sigma_{\text{H}}^{\uparrow}$ (see Fig.~\ref{fig: SigmaH}). 
This sign difference comes from $\hat{\Delta}^{\downarrow \downarrow}$ having a chirality opposite to that of $\hat{\Delta}^{\uparrow \uparrow}$. 
Also, under a relative sign change between $\Delta_{A_{1u}} $ and $\Delta_{A_{2u}}$, $\sigma_{\text{H}} \rightarrow - \sigma_{\text{H}}$, because
that relative sign determines which chirality of the pairing, $p_x -  i p_y$ (associated with spin-up)
or $p_x + i p_y$ (with spin-down), dominates. 
This can be seen from Fig.~\ref{fig: ImSigmaH0d8eV}, where we plot $\sigma_{\text{H}}(\omega)$ for a given frequency 
as a function of $4 \Delta_{A_{1u}} \Delta_{A_{2u}}/(|\Delta_{A_{1u}}| + |\Delta_{A_{2u}}|)^2$. 
Fig.~\ref{fig: ImSigmaH0d8eV} also shows that $\sigma_{\text{H}}=0$ if $\Delta_{A_{1u}}=0$ or $\Delta_{A_{2u}}=0$.
This is expected since in this case our mixed helical pairing state is reduced to one of the two single-representation helical states, either $A_{1u}$ or $A_{2u}$, for which the two chiral components completely compensate each other.  

\section{Estimation of the Kerr angle}
\begin{figure}[h]
\centering
\includegraphics[width=0.45\linewidth,trim={0 5mm 0 0}]{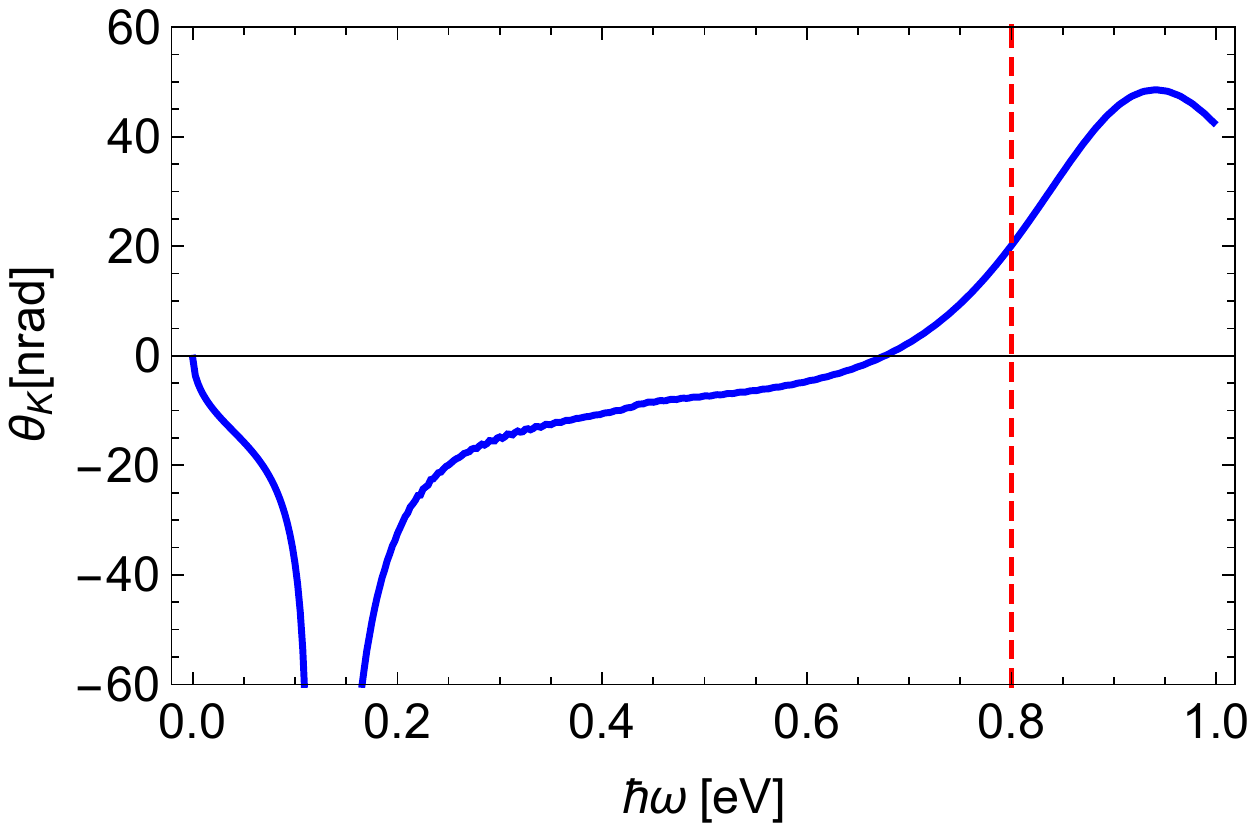}
\caption{(color online) Kerr angle as a function of $\hbar \omega$. Band parameters and the values of $\{\Delta_{A_{1u}}, 
\Delta_{A_{2u}} \}$ are the same as in Fig.~\ref{fig: SigmaH}.
}
\label{fig: Kerr}
\end{figure}

From the calculated $\sigma_{\text{H}}(\omega)$ we can estimate the Kerr angle for $\hbar \omega=0.8 \mathrm{eV}$ using
\begin{gather}
\theta_{\text{K}}(\omega) = \frac{4\pi}{\omega d} \mathrm{Im}[ \frac{ \sigma_{\text{H}}(\omega) } { n(n^2-1)}],
\label{eq: Kerr}
\end{gather}
where $n=n(\omega)$ is the complex index of refraction, given by
\begin{gather}
n=\sqrt{\epsilon_{ab}(\omega)}, \\
\epsilon_{ab}(\omega) = \epsilon_\infty+ \frac{4\pi i}{\omega} \sigma(\omega). 
\end{gather}
$\epsilon_{ab}(\omega)$ is the permeability tensor in the $ab$-plane.
$\epsilon_\infty=10$~\cite{Taylor:12} is the background permeability. 
$d=6.8 \AA$ is the inter-layer spacing along the $c$-axis. 
$\sigma(\omega)$ is the optical longitudinal conductivity. 
Following Ref.~\cite{Taylor:12} we use a simple Drude model for $\sigma(\omega)$
\begin{gather}
\sigma(\omega) = -  \frac{\omega_{\text{pl}}^2}{4\pi i (\omega + i \Gamma)},
\end{gather}
where $\omega_{\text{pl}}= 2.9 \, \mathrm{eV}$ is the the plasma frequency and $\Gamma=0.4 \, \mathrm{eV}$ is an elastic scattering rate.
At $\hbar \omega=0.8 \, \mathrm{eV}$, $\sigma(\omega) = 0.33 + i \, 0.67, \epsilon_{ab} = -0.52 + i \, 5.20$, and  $n=1.53 + i \, 1.69$.
Plugging this $n$ value into Eq.~\eqref{eq: Kerr} and using the $\sigma_{\text{H}}( \hbar\omega = 0.8 \text{eV})$ value from Fig.~\ref{fig: ImSigmaH0d8eV}
we get $ \theta_{\text{K}} \approx 20  \, \mathrm{nrad}$. 
The whole frequency dependence of $\theta_{\text{K}}$ is shown in Fig.~\ref{fig: Kerr}. 

\section{Results for chiral p-wave pairing}
For easy comparison, in this section, we reproduce some of the results of $\sigma_{\text{H}}$ and $\theta_{\text{K}}$ obtained in Ref.~\cite{Taylor:12} for the chiral $p_x+ i p_y$ pairing. 
In this case the pairing is between spin up and down with the nonzero order parameter components given by
\begin{gather}
\hat{\Delta}^{\uparrow \downarrow} (\vec{k})
=
\Delta_{0}
\begin{pmatrix}
 \sin k_x   &     0     \\
0       					&   i \sin k_y
\end{pmatrix}, 
\hat{\Delta}^{\downarrow \uparrow} (\vec{k}) =  \hat{\Delta}^{\uparrow \downarrow} (\vec{k}). 
\end{gather}
In the absence of SOC, the BdG Hamiltonian can be again decomposed into two decoupled blocks. One of them is 
\begin{gather}
\hat{H}_{\text{BdG}}^{\uparrow} (\vec{k})
= 
\begin{pmatrix}
\hat{H}_{\text{N}}(\vec{k})  &   \hat{\Delta}^{\uparrow\downarrow}(\vec{k})\\
[ \hat{\Delta}^{\uparrow\downarrow}]^\dagger(\vec{k})  & - [\hat{H}_{\text{N}}]^{T}(-\vec{k}) 
 \end{pmatrix},
 \label{eq: HBdG3}
\end{gather}
which is written in the basis $\{c^\dagger_{d_{xz},\uparrow} (\vec{k}), c^\dagger_{d_{yz},\uparrow}(\vec{k}), c_{d_{xz},\downarrow} (-\vec{k}), c_{d_{yz},\downarrow}(-\vec{k}) \}$. 
The other block is for the same basis but with opposite spin, and we denote it by $\hat{H}_{\text{BdG}}^{\downarrow} (\vec{k})$. Because $\hat{\Delta}^{\downarrow \uparrow} =  \hat{\Delta}^{\uparrow \downarrow}$,
$\hat{H}_{\text{BdG}}^{\downarrow}=\hat{H}_{\text{BdG}}^{\uparrow}$. 
Since the expression of $\hat{H}_{\text{BdG}}^{\uparrow} $ in Eq.~\eqref{eq: HBdG3} is almost identical to that in Eq.~\eqref{eq: HBdG}
we can immediately write down the expression of $\sigma_{\text{H}}^\uparrow$,
\begin{gather}
\frac{\sigma_{\text{H}}^{\uparrow}(\omega) }{e^2/\hbar}=  \sum_{\vec{k}}
 \frac{ \mathcal{F}_{\vec{k}} }{E_{+}(\vec{k})  E_{-}(\vec{k})  [E_{+}(\vec{k})  + E_{-}(\vec{k})]  [(E_{+}(\vec{k})  + E_{-}(\vec{k}) )^2 -( \omega+i\delta)^2]},
\end{gather}
where
\begin{subequations}
\begin{align}
\mathcal{F}_{\vec{k}} 
& =  - 32 \;  (t-\tilde{t}) (t^\prime)^2 \Delta_0^2  (\sin^2 k_x \cos k_y + \sin^2 k_y \cos k_x) \sin^2 k_x \sin^2 k_y \\
E_{\pm} & = \sqrt{ \frac{-\alpha \pm \sqrt{\alpha^2 - 4 \beta}}{2}}, \\
\alpha & = - (\xi_{xz}^2 +2  |\Delta_0|^2 + \xi_{yz}^2 + 2 \lambda^2), \\
\beta  & = (\xi_{xz}^2 +  |\Delta_0|^2)  (\xi_{yz}^2 +  |\Delta_0|^2) + \lambda^4  -2 \lambda^2 \xi_{xz} \xi_{yz}.
\end{align}
\end{subequations}
As a consequence of $\hat{H}_{\text{BdG}}^{\downarrow}=\hat{H}_{\text{BdG}}^{\uparrow}$, $\sigma_{\text{H}}^{\downarrow}=\sigma_{\text{H}}^{\uparrow}$
for the chiral p-wave pairing such that $\sigma_{\text{H}}= \sigma_{\text{H}}^{\downarrow}+ \sigma_{\text{H}}^{\uparrow}= 2 \sigma_{\text{H}}^{\uparrow}$. This should be contrasted with the previous $A_{1u}+ i A_{2u}$ pairing case where $\sigma_{\text{H}}^{\downarrow}$
and $\sigma_{\text{H}}^{\uparrow}$ carry opposite signs. 

\begin{figure}[htp]
\centering
\includegraphics[width=0.45\linewidth,trim={0 5mm 0 0}]{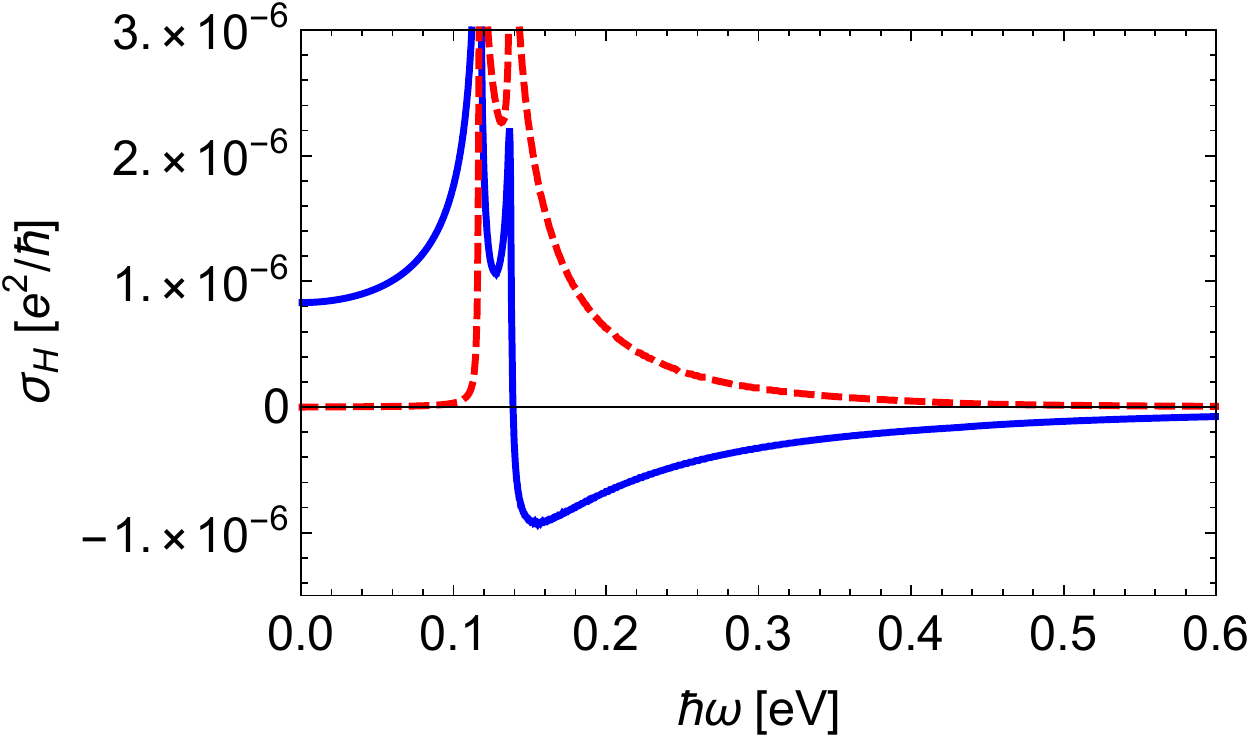}
\hspace{5mm}
\includegraphics[width=0.4\linewidth,trim={0 5mm 0 0}]{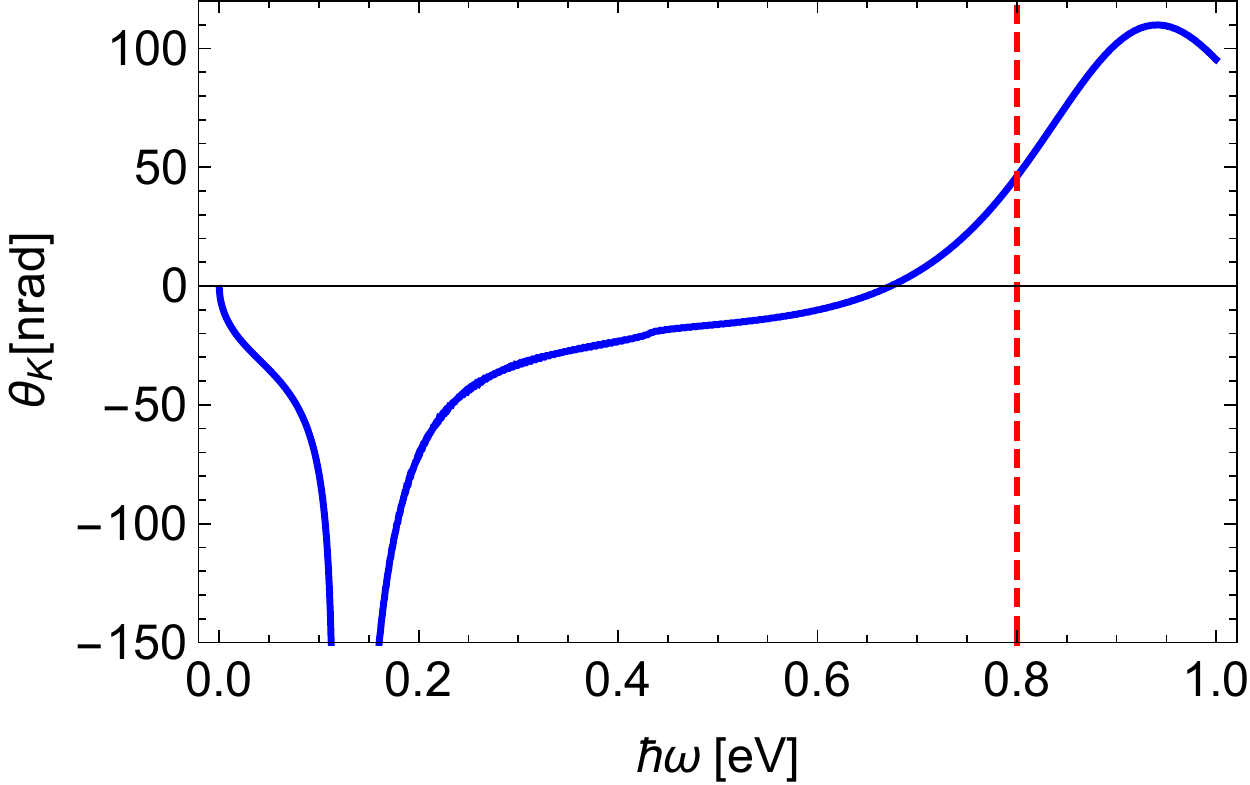}
\caption{Results of $\sigma_{\text{H}}$ and $\theta_{\text{K}}$ for the chiral p-wave pairing. For this calculation
the normal state band parameters used are the same as in Fig.~\ref{fig: SigmaH};  $\Delta_0=0.23 \, \mathrm{meV}$. 
In the left plot, the blue solid (red dashed) line is for the real (imaginary) part of $\sigma_{\text{H}}$. 
In the right plot, the dashed line shows that $\theta_{\text{K}}(\hbar \omega = 0.8 \, \mathrm{eV})\approx 46 \, \mathrm{nrad}$. 
}
\label{fig: chiralpnumerical}
\end{figure}

Fig.~\ref{fig: chiralpnumerical} shows the corresponding numerical results of $\sigma_{\text{H}}$ and $\theta_{\text{K}}$. Here, $\theta_{\text{K}}$ is evaluated from $\sigma_{\text{H}}$ using Eq.~\eqref{eq: Kerr}
with the same index of refraction $n(\omega)$. From Fig.~\ref{fig: chiralpnumerical} we see that $\theta_{\text{K}}(\hbar \omega = 0.8 \, \mathrm{eV})\approx 46 \, \mathrm{nrad}$, which is about twice larger than
the estimated value for the $A_{1u}+ i A_{2u}$ pairing (see Fig.~\ref{fig: Kerr}). The larger $\theta_{\text{K}}$ in the current case comes from the absence of cancellation, i. e. $\sigma_{\text{H}}^{\downarrow}$
and $\sigma_{\text{H}}^{\uparrow}$ carry the same sign and the two add together instead of cancel each other.  

\section{Absence of cross-gradient terms in the Ginzburg-Landau free energy}
Using the $A_{1u}+iA_{2u}$ state as an example, we argue that there are no cross-gradient terms in the free energy that are quadratic in the order parameter components. There are in total four possible such terms that involve both order parameter components as well as gradients in both $x$ and $y$ directions:
 \begin{equation}
 a\partial_x\Delta^\ast_{A_{1u}}\partial_y\Delta_{A_{2u}} + b\partial_y\Delta^\ast_{A_{1u}}\partial_x\Delta_{A_{2u}}+ c\partial_x\Delta^\ast_{A_{2u}}\partial_y\Delta_{A_{1u}} + d\partial_y\Delta^\ast_{A_{2u}}\partial_x\Delta_{A_{1u}}
 \end{equation}
where the coefficients $a,b,c$ and $d$ assume the values of 1 or -1 and we have dropped an overall Ginzburg-Landau coefficient. The general requirement is that the free energy be real and invariant under all $D_{4h}$ point group symmetry transformations. Bearing in mind the symmetry properties of the two individual components, one can check that no single set of $\{a,b,c,d\}$ ensures the invariance of the above free energy under all symmetry operations. One thus concludes that these cross-gradient terms must be absent. Same argument applies to the $B_{1u}+iB_{2u}$ state.
